\documentclass[%
 reprint,
superscriptaddress,
nofootinbib,
 amsmath,amssymb,
 aps,
nolongbibliography
]{revtex4-2}
\usepackage{amssymb}
\usepackage{bm}
\usepackage{dcolumn}
\usepackage{graphicx}
\usepackage[dvipsnames, usenames]{xcolor}
\definecolor{linkcolor}{rgb}{0.0,0.3,0.5}
\definecolor{myNavy}{rgb}{0,0,.5} 
\usepackage[breaklinks=true]{hyperref}
\hypersetup{
    colorlinks=true,
    linkcolor=linkcolor,
    filecolor=linkcolor,
    urlcolor=linkcolor,
    citecolor=linkcolor
}
\usepackage{placeins}
\usepackage[normalem]{ulem}
\usepackage{orcidlink}
\newcommand{\eqbreak}{\nonumber\\&~}
\newcommand{\neweq}{\nonumber\\}
\newcommand{\popu}{{\mathrm{pop}}}
\newcommand{\pe}{{\mathrm{pe}}}

\newcommand{\ml}{{\mathrm{ml}}}
\newcommand{\hi}{{\mathrm{hi}}}

\newcommand{\boldtheta}{{\boldsymbol{\theta}}}
\newcommand{\boldmu}{{\boldsymbol{\mu}}}
\newcommand{\boldLambda}{{\boldsymbol{\Lambda}}}
\newcommand{\boldx}{{\boldsymbol{x}}}

\newcommand{\sci}[1]{\text{$ \times 10^{ #1}$}}

\begin{document}

\newcommand{\milan}{\affiliation{Dipartimento di Fisica ``G. Occhialini'', Universit\'a degli Studi di Milano-Bicocca, Piazza della Scienza 3, 20126 Milano, Italy}}
\newcommand{\infn}{\affiliation{INFN, Sezione di Milano-Bicocca, Piazza della Scienza 3, 20126 Milano, Italy}}

\title{
Population-level correlations in Bayesian statistics: \\ An illustrative model for gravitational-wave astronomy 
}
\author{Caroline B. Owen$\,$\orcidlink{0000-0002-6800-7132}}%
\email{cbo2@williams.edu}
\milan \infn

\author{Alexandre Toubiana$\,$\orcidlink{0000-0002-2685-1538}}%
\milan \infn
\author{Davide Gerosa$\,$\orcidlink{0000-0002-0933-3579}}
\milan \infn

%\date{\today}

\begin{abstract}
With increasingly large numbers of gravitational-wave events, population inference is now beginning to move beyond predictions of marginal distributions and to probe correlations between compact-binary parameters such as masses, spins, and redshifts.
These correlations have strong constraining power for both astrophysics and tests of general relativity.
In this paper, we present an idealized analytical model to study the interplay between single-event correlations, systematic biases, and population-level correlations.
With this, we investigate the potential emergence of false-positive measurements of population-level correlations.
We quantify how the presence of correlations at the single-event level between a pair of parameters increases the uncertainty of population-level correlations for those parameters, potentially obscuring the true underlying population correlation if present.
We also find that if waveform systematics lead to biases that are correlated across the catalog (which is likely, because certain regions of the parameter space are more difficult to model), this can be effectively absorbed by a population analysis that targets correlations and can be misinterpreted as such.
This simple Gaussian-based model may serve as a broad compass for future, more detailed explorations.
\end{abstract}

\maketitle

\section{Introduction}\label{SEC:Intro}
Population analyses of gravitational-wave (GW) sources, which combine multiple events within the hierarchical Bayesian framework~\cite{Mandel:2018mve,Vitale:2020aaz}, are becoming increasingly common tools to unveil the formation mechanisms of compact binaries, %~\cite{Talbot:2018cva,Fishbach:2018edt,LIGOScientific:2025pvj,Antonini:2025ilj,Adamcewicz:2025phm,Tong:2025wpz,Tenorio:2025nyt,Banagiri:2025dmy,Guttman:2025jkv,Ray:2025xti,Tiwari:2025oah,Tong:2025xir,Gennari:2025nho,Heinzel:2024jlc,Cheng:2026bpc,LIGOScientific:2026ctl}, \dg{there are a billion papers here, and it's impossibile to cite everyone. This selection appears ad hoc and kind of biased as well, with several papers from the same group and other groups ignored. But it's an impossible task. Let's remove all these citations (both here and below for tests of GR and cosmology), and cite reviews instead}
improve our ability to test general relativity, %~\cite{Payne:2023kwj,Magee:2023muf,Zhong:2024pwb,LIGOScientific:2026qni,LIGOScientific:2026fcf,LIGOScientific:2026wpt}, 
and measure cosmological parameters
%~\cite{Mastrogiovanni:2021wsd,LIGOScientific:2025jau,MaganaHernandez:2025cnu,Gennari:2026dfy}. 
(e.g. \cite{LIGOScientific:2026ctl,LIGOScientific:2026uyd,LIGOScientific:2026qni}).
As the size of GW catalog increases, we become able not only to infer the population-level marginal distributions of binary parameters, but also to characterize the correlations between them \cite{Callister:2024cdx,Heinzel:2024jlc,Antonini:2025ilj,Tenorio:2025nyt,Banagiri:2025dmy,Guttman:2025jkv,Ray:2025xti,Tiwari:2025oah,Tong:2025xir,Gennari:2025nho,Alvarez-Lopez:2025ltt,Cheng:2026bpc}.
This, in turn, enhances our ability to distinguish between different formation scenarios~\cite{Mandel:2018hfr,Mapelli:2021taw} and to identify the origin of possible deviations from general relativity~\cite{Zhong:2024pwb,Payne:2024yhk}. 
Given the central role that measurements of population-level correlations are likely to play in shaping our understanding of the Universe, it is essential to understand the factors that may affect them.

A negative correlation between the mean of the effective-spin distribution and the mass ratio was initially identified in the GWTC-2.1 catalog~\cite{Callister:2021fpo} and further confirmed with subsequent data \cite{KAGRA:2021duu}.
Interestingly, as more events have been added to the catalog, the evidence for this correlation weakened, while evidence for a correlation between the width of the effective-spin distribution and mass ratio has increased~\cite{LIGOScientific:2025pvj,LIGOScientific:2026ctl}.

The initially inferred trend follows the same direction as the well-known degeneracy between effective spin and mass ratio at the single-event parameter-estimation level~\cite{Ng:2018neg}. 
Reference~\cite{Callister:2021fpo} performed additional analyses to strengthen the interpretation that this single-event degeneracy was not responsible for the inferred population-level correlation.
Indeed, hierarchical analyses are unbiased in the ideal limit: if the model used to analyze the data is correct, they should, on average, recover the correct population properties.
However, whether correlations at the individual-event level can facilitate false measurements of population-level correlations, 
especially due to statistical fluctuations in the finite-number-of-events regime, remains an important question.

In addition to such statistical effects, systematic effects may also contribute to false measurements of population-level correlations. 
At the single-event level, the finite accuracy of the waveform models used in parameter estimation of individual events can lead to systematic biases in the inferred source properties~\cite{Owen:2023mid,Dhani:2024jja,Kapil:2024zdn,Chandramouli:2024vhw,LIGOScientific:2025rsn,LIGOScientific:2025cmm}.
Since these biases are source-dependent, they could in turn mimic spurious correlations at the population level.
Additionally, population analyses require choosing a model for the distribution of events.
If this model is mispecified, the reconstruction of the population properties can be compromised~\cite{Romero-Shaw:2022ctb,Cheng:2023ddt,Miller:2026buq,Mould:2026sww}, including the inferred correlations between binary parameters.  

In this paper, we systematically investigate false-positive measurements of population-level correlations when no such correlations are present in the true population. 
We consider three possible causes: 
(i) statistical fluctuations, including the role of correlations between parameters in individual observations; 
(ii) systematic biases in single-event measurements; 
and (iii) systematic biases arising from an incorrect choice of population prior.
We construct a simple toy model to explore how these factors affect the measurement of population-level correlations between pairs of binary parameters. 
While considerably simpler than the approaches commonly used to infer the astrophysical properties of compact-binary populations, our framework closely resembles the approach developed in Ref.~\cite{Zhong:2024pwb} to constrain multidimensional deviations from general relativity.
As a result, our findings are particularly relevant to and directly translatable to that science case, while also providing insight into qualitative behaviors in other inference setups.

The remainder of this paper is organized as follows. In Sec. \ref{SEC:ToyModel}, we set up our toy model and demonstrate how it models statistical and systematic errors at the single-event and population levels.
In Sec. \ref{SEC:ProbDists}, we explore how statistical fluctuations in finite catalogs can impact the probability of obtaining a credible false-positive measurement of correlation.
Finally, in Sec. \ref{SEC:Conclusion} we summarize the results and discuss their implications. 
% %

%%%%%%%%%%%%%%%%%%%%%%%%%%%%%%%%%%%%%%%%%%%%%%%%%%%%%%%%%%%%%%%%%%%%%%%%%%%%%%%%%%%%
\section{Model setup}\label{SEC:ToyModel}
%-----------------------------------------------------------------------------------
We construct a simple toy model to examine correlations between two binary parameters $\boldtheta = (\theta_1,\theta_2)$.
We assume that we have observed a catalog $\{d_k\}$ of $N$ signals with $k\in[1,...,N]$.
Each signal $d_k= n_k + h^{\star}(\boldtheta^{\star}_{k})$ is a sum of detector noise $n_k$ and the true\footnote{Throughout this manuscript, we will use the symbol $\star$ to indicate true values of the single-event parameters $\boldtheta^\star$, population parameters $\boldLambda^\star$, waveform $h^\star$, and population distibution $\pi^\star$.} 
 waveform $h^{\star}$ produced by true source parameters  $\boldtheta^{\star}_{k}$.
The source parameters of each signal are drawn from some underlying population model $\pi^{\star}(\boldtheta|\boldLambda^{\star})$ with population parameters~$\boldLambda^{\star}$.

We assume that both the single-event and population likelihoods are well described by multivariate Gaussian distributions. Additionally, we employ a multivariate Gaussian model to recover the population properties. 
For completeness, we recall the expression of the multivariate Gaussian distribution.  
Consider a random vector $\boldx$ of dimension $\ell$
distributed according to a multivariate Gaussian distribution with mean vector $\boldmu$ and covariance matrix $\Sigma$.
The probability distribution function (PDF)  of $\boldx$ is 
\begin{align}
    \mathcal{N}(\boldx|\boldmu,\Sigma) = \frac{\exp\left[-\frac{1}{2}(\boldx-\boldmu)^{T}\Sigma^{-1}(\boldx-\boldmu)\right]}{\sqrt{(2\pi)^{\ell}|\Sigma|}}\,.
\end{align}

The covariance matrix is a symmetric matrix where the diagonal components $\Sigma_{ii} =\sigma_i^2$ are the variances of the marginal distributions and the off-diagonal components $\Sigma_{ij} =\rho_{ij}\sigma_i\sigma_i$ are proportional to Pearson's correlation coefficients $\rho_{ij} \in (-1,1)$. 
These coefficients quantify linear correlations such that when $\rho_{ij}$ is positive (negative), $x_i$ and $x_j$ are (anti)correlated, and when $\rho_{ij}=0$, they are uncorrelated. 
The components of the inverse covariance matrix can be obtained by inverting the Fisher information matrix~\cite{Vallisneri:2007ev}
\begin{align}\label{eq:fim}
   \left(\Sigma^{-1}\right)_{ij} = -\frac{\partial ^2}{\partial x ^i\partial x ^j} \log \mathcal{N}(\boldx|\boldmu,\Sigma)\Bigg|_{\boldx=\boldmu} \,.
\end{align}
Typically, in GW astronomy, we quote 90\% credible intervals on the marginal distributions of a given posterior.
For a multivariate Gaussian, 90\% of the probability density of the marginal distribution on $x_i$ is contained within an interval of ${2\sqrt{2}\sigma_{i}\mathrm{Erf}^{-1}(0.9)} $ centered about $\mu_i$. 

%-----------------------------------------------------------------------------------
\subsection{Single-event parameter estimation}\label{SUBSEC:BinaryPE}

Assuming the detector noise to be zero-mean and Gaussian distributed, the single-event likelihood is
\begin{align}\label{EQ:gaussianlikeli}
\mathcal{L}(d|\boldtheta) \propto 
\exp\left[-\frac{1}{2}\left(d-h(\boldtheta)\middle|d-h(\boldtheta)\right)\right]\,,
\end{align}
where $(\cdot|\cdot)$ denotes the inner product weighted by the one-sided noise power spectral density and $h(\boldtheta)$ is the chosen waveform approximant. 

In the large signal-to-noise ratio (SNR) limit, this likelihood is well described by a multivariate Gaussian distribution on the single-event parameters  
\begin{align}\label{EQ:GaussDists1}
\mathcal{L}(d|\boldtheta) \propto \mathcal{N}(\boldtheta|\boldtheta_\ml,\Sigma_\pe)\,, 
\end{align}
centered around the maximum likelihood values $\boldtheta_\ml$ and with a covariance matrix $\Sigma_\pe$. 

In this toy model, we prescribe $\Sigma_\pe$ via standard deviations $\sigma_{\pe,1}$ and $\sigma_{\pe,2}$ and correlation coefficient $\rho_\pe$.
The $\sigma_{\pe,i}$ are a surrogate for SNR, with which they scale inversely in the large-SNR limit for a given location in parameter space and model $h(\boldtheta)$.
The coefficient $\rho_\pe$ captures the correlation between $\theta_1$ and $\theta_2$ in the single-event likelihood that arises due to the functional form of $h(\boldtheta)$. 
In this simplified approach, the maximum-likelihood point is a sufficient statistic for describing the data, so in the following we will refer to the data $d_k$ through $\boldtheta_{\ml,k}$.

The maximum likelihood point for a given event can be shifted from the true value for two reasons: (i) the presence of noise in the detector, which causes a statistical shift, and (ii) waveform mismodeling, which causes a systematic bias. Defining  
\begin{align}\label{EQ:bias_def}
    \Delta\boldtheta = \boldtheta_\mathrm{ml} - \boldtheta^{\star}\,,
\end{align} 
we can write
\begin{align}
    \Delta\boldtheta = \Delta \boldtheta_\mathrm{stat}+ \Delta \boldtheta_\mathrm{sys}\,.
\end{align}
When the shift is small, and under the Gaussian likelihood approximation, these contributions can be approximated as~\cite{Cutler:2007mi,Chandramouli:2024vhw}
\begin{align}\label{EQ:deltastat}
    \Delta \boldtheta_\mathrm{stat} =&~\Sigma_{\pe}\cdot\left(\frac{\partial h}{\partial\boldtheta} \bigg |n\right)\bigg|_{\boldtheta=\boldtheta^{\star}} \,,
\\\label{EQ:deltasys}
    \Delta \boldtheta_\mathrm{sys} =&~ \Sigma_{\pe}\cdot\left(\frac{\partial h}{\partial\boldtheta} \bigg |h -h^{\star}\right)\bigg|_{\boldtheta=\boldtheta^{\star}}  \,.
\end{align}
With the likelihood model of Eq.~\eqref{EQ:gaussianlikeli}, $\Delta \boldtheta_\mathrm{stat}$ is distributed according to a normal distribution with zero mean and covariance matrix $\Sigma_\pe$. 
The difference between the waveform approximant and the true waveform $h -h^{\star}$, and consequently $\Delta \boldtheta_\mathrm{sys}$,  will have some particular dependence on $\boldtheta^{\star}$ and potentially additional binary parameters.
In the following Section, we consider several illustrative examples.

Thus, our mock catalog $\{\boldtheta_{\ml,k}\}$ is constructed as follows:
\begin{enumerate}
    \item Draw $N$ samples $\boldtheta^\star$ from the chosen population distribution 
    $\pi^\star(\boldtheta|\Lambda^\star)$.
    
    \item For each $\boldtheta^\star$, we draw $\Delta \boldtheta_{\rm stat}$ from the Gaussian distribution
    $\mathcal{N}(\boldsymbol{0},\Sigma_{\pe})$. 
    If systematic effects are included, we additionally compute 
    $\Delta \boldtheta_{\rm sys}$ according to the chosen prescription. Finally, $\boldtheta_{\ml}=\boldtheta^\star+\Delta \boldtheta$.  
    
\end{enumerate}

\subsection{Hierarchical inference}\label{SUBSEC:HI}
Once the mock catalog of observations is generated, the next step is hierarchical inference to recover the population parameters $\boldLambda$. 
Regardless of the underlying distribution the events are drawn from, we model the population with the multivariate Gaussian distribution 
\begin{align}\label{EQ:GaussDists2}
\pi(\boldtheta|\boldLambda)= \mathcal{N}(\boldtheta|\boldmu_\popu,\Sigma_\popu)\,.
\end{align}
with population parameters 
\begin{align}
    \mathbf{\Lambda}= (\mu _{\popu,1},\mu _{\popu,2},\sigma _{\popu,1},\sigma_{\popu,2},\rho_\popu)\,.
\end{align} 
The coefficient $\rho_\popu$ is the population parameter that quantifies the population-level correlation between $\theta_1$ and $\theta_2$.
It is the measurement of this parameter that interests us here.

We take the population likelihood of observing the catalog $\{\boldtheta_{\ml,k}\}$ given $\boldLambda$ to be the product of the single-event likelihoods, marginalized over the population model
\begin{align}\label{EQ:poplike1}
    \mathcal{L}(\{\boldtheta_{\ml,k}\}|\boldLambda) =&~ \prod_{k}^{N}\int d\boldtheta_k \mathcal{L}(\boldtheta_{\ml,k}|\boldtheta_k)\pi(\boldtheta_k|\boldLambda)\,.
\end{align}
This is a simplified likelihood that has been marginalized over the rate of binary mergers and that disregards selection effects.
Note that selection effects are also not usually included in hierarchical tests of general relativity~\cite{LIGOScientific:2026qni,LIGOScientific:2026fcf,LIGOScientific:2026wpt}, although frameworks for doing so do exist~\cite{Payne:2023kwj}. 

In our setup, where both the population model $\pi(\boldtheta|\boldLambda)$ and parameter estimation likelihoods $ \mathcal{L}(\boldtheta_{\ml}|\boldtheta)$ are distributed according to multivariate Gaussian distributions, the above integrals can be done analytically 
\begin{align}\label{EQ:simplepoplike}
    \mathcal{L}(\{\boldtheta_{\ml,k}\}|\mathbf{\Lambda}) = \prod_{k=1}^N \mathcal{N}(\boldtheta_{\ml,k}|\boldmu_\popu,\Sigma_\popu+\Sigma_\pe)\,.
\end{align}

The population likelihood will be maximized by $\mathbf{\Lambda}_{\ml}$  which is specified by $\boldsymbol{\mu}_\ml = \boldsymbol{\hat\mu}$ and $\Sigma_\ml = \hat{\Sigma} - \Sigma_\pe$. The hatted quantities are the empirical mean and covariance matrix of the set of $N$ observations $\{\boldtheta_{\ml,k}\}$.
In this model, both $\boldsymbol{\hat\mu}$ and $\hat\Sigma$ can be computed algebraically
\begin{align}\label{EQ:max_ml}
    \boldsymbol{\hat\mu}  =& \frac{1}{N}\sum_{k =1}^N \boldtheta^k_{\mathrm{ml}}\,,\neweq
   \hat{\Sigma}  =& \frac{1}{N}\sum_{k =1}^N (\boldtheta^k_{\mathrm{ml}} -\boldsymbol{\hat\mu})^T(\boldtheta^k_{\mathrm{ml}} -\boldsymbol{\hat\mu})\,,
\end{align}
allowing us to determine $\rho_{\popu,\ml}$ directly. Note that the population likelihood is maximized by the biased estimator of the covariance matrix, rather than by the unbiased estimator, which uses a factor of $1/(N-1)$ instead of $1/N$. Nevertheless, this estimator is unbiased in the limit $N \to +\infty$.

In the absence of mismodeling in both waveform approximant and population model, the maximum likelihood values $\{\boldtheta_{\ml,k}\}$ are generated by drawing the source parameters $\{\boldtheta^\star_k\}$ from the Gaussian distribution $\mathcal{N}(\boldtheta|\boldmu^\star_{\popu},\Sigma^{\star}_{\popu})$ and shifting them by $\{\Delta \boldtheta_{\mathrm{stat},k}\}$ drawn from  $\mathcal{N}(\boldtheta|\boldsymbol{0},\Sigma_\pe)$.
In this case, $N \hat{\Sigma}$ is distributed according to a Wishart distribution with $N -1$ degrees of freedom and with scale matrix $ \Sigma^{\star}_{\popu} + \Sigma_\pe$.
In general, the expectation value of a function of random variables is not equal to that function evaluated at the expectation values of those random variables.
While, it is straightforward to show that $\langle\Sigma_\ml \rangle =\Sigma^{\star}_{\popu}$, $\rho_{\popu,\ml}$ is a nontrivial function of the components of $\hat{\Sigma}$ and $\Sigma_\pe$, and it is not necessarily the case that $\langle\rho_{\popu,\ml} \rangle = \rho^\star_{\popu}$.
While we are not aware of an analytic expression for this expectation value, it will, in general, depend on $N$ and the components of $\Sigma_\pe$, including $\rho_\pe$.
However, as $N\rightarrow\infty$, the Wishart distribution becomes more strongly peaked at its expectation value and thus the peak of the distribution on $\rho_{\popu,\ml}$ is pushed toward $\rho^\star_{\popu}$.

If systematic error is present at either the single-event or population level, $N \hat{\Sigma}$ will no longer be governed by the Wishart distribution.
The distribution that $N \hat{\Sigma}$, and consequently $\rho_{\popu,\ml}$, takes depends on the specific form of the mismodeling at hand.
For example, if $\{\Delta \boldtheta_{\mathrm{sys},k}\}$ depends on the source parameters such a way as to induce additional correlation in  $\{\boldtheta_{{\ml},k}\}$, $\rho_{\popu,\ml}$ may be biased from $\rho^\star_{\popu}$.

In general, mismodeling at the population inference level can also lead to bias in the recovered population parameters.
However, it is here that the simplicity of our toy model becomes a shortcoming.
The empirical covariance matrix $\hat{\Sigma}$ as given in Eq. \eqref{EQ:max_ml} will capture the correlation in $\{\boldtheta_{\ml,k}\}$ regardless of the distribution from which $\{\boldtheta^\star_{k}\}$ is drawn.
If the true population distribution of the single-event parameters is uncorrelated, then we expect $\rho_{\popu,\ml}$ to tend to 0 in the large $N$ limit even if the Gaussian population model is not overall a good representation of the true distribution. 

\subsection{Quantifying false measurements}\label{SUBSEC:Quantifying}
We restrict our consideration to underlying population models $\pi^{\star}(\boldtheta|\boldLambda^{\star})$ in which $\theta_1$ and  $\theta_2$ are uncorrelated at the population level, i.e. $\rho^{\star}_\popu =0$, because we seek to understand the circumstances in which population-level correlation may be measured despite such correlation not being present in the true distribution.  

As with the single-event parameters,  $\rho_{\popu,\ml}$ can be shifted from $\rho^{\star}_\popu$ due to both statistical and systematic effects.
However, whether a given catalog yields a credible measurement of population-level correlation will depend not only on $\rho_{\popu,\ml}$ but also on the uncertainty of the measurement.   
To estimate the credible intervals, we make the additional assumption that the population likelihood is well described as a five-dimensional multivariate Gaussian distribution  
\begin{align}
\mathcal{L}(\{\boldtheta_{\ml,k}\}|\boldLambda) \propto \mathcal{N}(\boldLambda|,\mathbf{\Lambda}_{\ml}, \Sigma_{\hi})
\end{align}
centered about $\mathbf{\Lambda}_{\ml}$. Following Eq.~\eqref{eq:fim}, the inverse of the covariance matrix of the likelihood is
\begin{align}\label{EQ:FIMhi}
   (\Sigma_{\hi}^{-1})_{ij}=-\frac{\partial ^2}{\partial \Lambda ^i\partial \Lambda ^j}\log\mathcal{L}(\{\boldtheta_{\ml,k}\}|\boldLambda)\Bigg|_{\boldLambda=\boldLambda_\ml}\,. 
\end{align}
where $\mathcal{L}(\{\boldtheta_{\ml,k}\}|\boldLambda)$ is given by Eq~\eqref{EQ:simplepoplike} \cite{Gair:2022fsj,DeRenzis:2024dvx}. 

We take the prior distribution on the population parameters to be uniform\footnote{The uniform prior on $\rho_\popu$ is an instance of a Lewandowski-Kurowicka-Joe (LKJ) distribution with shape parameter 1. The LKJ distribution is a probability distribution on correlation matrices and can be tuned to disfavor (favor) correlations by selecting a shape factor greater (less) than~1~\cite{LEWANDOWSKI20091989}. While not investigated in this study, it is reasonable to suspect that increasing (decreasing) the shape factor would decrease (increase) the probability of a false-positive identification of correlation.} and sufficiently wide such that the standard deviation ${\sigma_\rho}_\popu$ of the marginalized posterior on $\rho_\popu$  can then be extracted from $\Sigma_{\hi}$.
For several realizations of the data $\{\boldtheta_{\ml,k}\}$, we employed an Markov Chain Monte Carlo algorithm to sample the true posterior and have verified that ${\sigma_\rho}_\popu$ as computed above is a good approximation of that of the true posterior for the cases considered in this study, generally with a relative error of $\sim 1\%$.
As with $\rho_{\popu,\ml}$, the value of ${\sigma_\rho}_\popu$ for a given catalog will be impacted by statistical fluctuations and systematic effects. 

We consider a measurement to be consistent with {no} correlation if $\rho_{\popu} = 0$ falls within the 90\% credible interval of the marginalized posterior.
When this occurs, the ratio   
\begin{align}
    r_\rho = \frac{|\rho_{\popu,\ml}|}{\sqrt{2}\sigma_{\rho_\popu}\mathrm{Erf}^{-1}(0.9)}
\end{align}
is less than 1. 
To understand the impact of statistical fluctuations on the measurement of population-level correlation, we set out to estimate probability distributions $p(\rho_{\popu,\ml})$, $p(\sigma_{\rho_\popu})$, and $p(r_\rho)$ under various conditions. 

The toy model we have laid out allows us to account for statistical uncertainty and both statistical and systematic shifts in measuring the single-event and population parameters.% 
We investigate the effects of statistical uncertainty at the level of single-event parameter estimation by varying the standard deviations $\sigma_{\pe,1}$ and $\sigma_{\pe,2}$.
Varying  $\sigma_{\pe,1}$ and $\sigma_{\pe,2}$ will also impact the statistical shifts, the magnitude of which will tend to be larger when $\sigma_{\pe,1}$ and $\sigma_{\pe,2}$  are larger.
The correlation coefficient $\rho_\pe$ allows us to model the effects of correlation in the measurement of binary parameters for individual events.
We can explore the effects of various types of systematic error at the parameter estimation level by modeling $\Delta \boldtheta_{\mathrm{sys}}$ in different ways.

Varying the catalog size $N$ allows us to explore the effects of statistical uncertainty and shifts at the population level.
As the catalog size grows, the statistical uncertainty in measurements of population parameters decreases.
Similarly, the effects of catalog variance, in which the properties of the finite catalog differ from those of the underlying distribution, will decrease, thereby reducing statistical shifts.
We can explore the effects of population mismodeling by choosing an underlying population model $\pi^{\star}(\boldtheta|\boldLambda^{\star})$ that does not match the recovery model.

Our procedure is as follows:
We specify the underlying population distribution, $\pi^{\star}(\boldtheta|\boldLambda^{\star})$, the catalog size $N$, and the single-event likelihood $\sigma_{\pe,1}$, $\sigma_{\pe,2}$, $\rho_\pe$, $\Delta \boldtheta_\mathrm{sys}$. A single catalog consists of $N$ sets of source parameters $\{\boldtheta^{\star}_{k}\}$ drawn from $\pi^{\star}(\boldtheta|\boldLambda^{\star})$.
We determine the maximum likelihood values of the single-event parameters $\{\boldtheta_{\ml,k}\}$ based on the chosen single-event likelihood.
Finally we compute $\rho_{\popu,\ml}$, $\sigma_{\rho_\popu}$, and $r_\rho$ for the catalog. 
For a given configuration we perform this calculation $2\sci{5}$ times, allowing us to approximate the probabilty distributions $p(\rho_{\popu,\ml})$, $p(\sigma_{\rho_\popu})$, and $p(r_\rho$).
By computing the fraction of catalogs for which $r_\rho>1$, we can determine the probability that hierarchical inference will imply a population-level correlation between $\theta_1$ and $\theta_2$ despite no correlation being present in the true underlying population.

%%%%%%%%%%%%%%%%%%%%%%%%%%%%%%%%%%%%%%%%%%%%%%%%%%%%%%%%%%%%%%%%%%%%%%%%%%%%%%%%%%%%
\section{Results}\label{SEC:ProbDists}

%-----------------------------------------------------------------------------------
\subsection{No mismodeling}\label{SUBSEC:NoMismodeling}
We begin by considering the case where there is no mismodeling, either at the single-event parameter estimation level, i.e., $\Delta \boldtheta_\mathrm{sys} =0$, or at the population hierarchical inference level, i.e.,  $ \pi^{\star}(\boldtheta|\boldLambda^{\star}) = \mathcal{N}(\boldtheta|\mu^{\star}_{\popu}, \Sigma^{\star}_{\popu})$. 
Specifically, we consider an uncorrelated population distribution with $\mu^{\star}_{\popu,1}=\mu^{\star}_{\popu,2} = 0$,  $\sigma^{\star}_{\popu,1}=\sigma^{\star}_{\popu,2} = 1$ and $\rho^{\star}_{\popu} =0$. 
With no mismodeling present, we investigate how statistical effects alone impact the measurement of $\rho_{\popu}$.

We consider a grid of 15 single-event likelihoods 
with $\sigma_\pe = \sigma_{\pe,1} = \sigma_{\pe,2} \in \{0.1, 0.5,  0.9\}$ and $\rho_\pe \in \{-0.9,-0.5, 0.0,0.5,0.9\}$.
For each configuration, we consider catalogs with $N\in\{50,100\}$ events each, comparable to the current GW catalogs, % used in the LVK analyses~\cite{LIGOScientific:2025pvj,LIGOScientific:2026qni,LIGOScientific:2026fcf,LIGOScientific:2026wpt}, \dg{now we're at 200, let's avoid citations}
and approximate the distributions $p(\rho_{\popu,\ml})$ and $p(\sigma_{\rho_\popu})$ in the manner described above.  

\begin{figure*}
    \centering
    \includegraphics[width=0.9\textwidth]{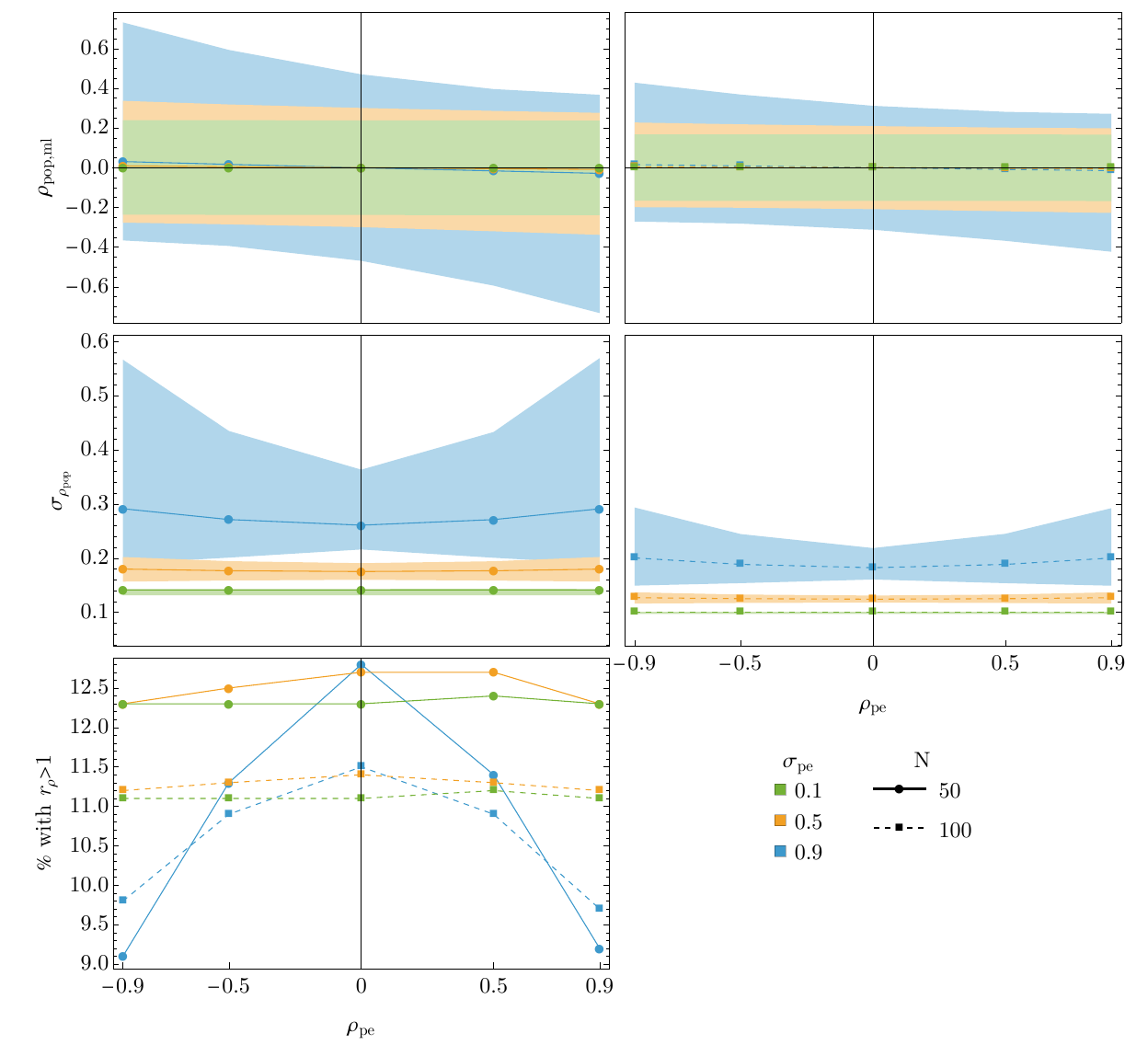}
    \caption{\label{FIG:grid}
    Top four panels show the median of the distributions $p(\rho_{\popu,\ml})$  and $p(\sigma_{\rho_\popu})$ as a function of $\rho_\pe$ for various values of $\sigma_\pe$ (colors) and $N$ (left vs. right; linestyles) in the absence of mismodeling at the single event or population levels. The $5\%$ and $95\%$ quantiles of the probability distributions enclose the shaded regions. The bottom panel shows the percentage of catalogs that produce a credible measurement of $\rho_\popu\neq0$, corresponding to $r_\rho>1$. The true population correlation is $\rho_\popu^{\star} =0$. We note two counterintuitive results: (1) There is a slight negative correlation between $\rho_{\pe}$ and $\rho_{\popu,\ml}$ such that when $\rho_{\pe}$ is positive (negative), there is a greater probability for values of $\rho_{\popu,\ml}$ that are more negative (positive), indicating that spurious population-level correlations are unlikely to be aligned with correlations in the single-event posteriors. (2) Because large values of  $\rho_{\pe}$ tend to increase the uncertainty in the measurement of $\rho_\popu$ as quantified by $\sigma_{\rho_\popu}$, larger single-event correlations can decrease the probability of a false positive measurement of $\rho_\popu\neq0$.
    }
\end{figure*}

The medians of these distributions are shown as functions of $\rho_\pe$ in the top four panels of Fig. \ref{FIG:grid}. The $5\%$ and $95\%$ quantiles of the distributions enclose the shaded regions.
The bottom panel plots the percentage of catalogs for which $r_\rho>1$. 

We first consider the cases where $\rho_\pe =0$. We see that even in the absence of single-event correlation, there is a nonzero probability of measuring $\rho_\popu\neq 0$.
However, the chance of a false-positive correlation measurement tends to decrease as both the precision of the single-event parameter estimation and the catalog size increase.   

Now, turning to the full set of configurations, we find that the median of $p(\rho_{\popu,\ml})$ is approximately zero for all configurations, as predicted for large catalogs in our above.
Although the absolute value of the median does increase slightly with both $|\rho_\pe|$ and $\sigma_\pe$, and decrease with $N$.
On the other hand, the spread of $p(\rho_{\popu,\ml})$ varies much more apparently among configurations, in particular increasing with $\sigma_\pe$ and decreasing with $N$.
Interestingly, while the shape of the distribution is much less impacted by $\rho_\pe$, there is a slight negative correlation between $\rho_{\pe}$ and $\rho_{\popu,\ml}$.
When $\rho_{\pe}$ is positive (negative), there is a greater probability for values of $\rho_{\popu,\ml}$ that are more negative (positive).
This negative correlation results from our choice to quantify the correlation using the dimensionless coefficient $\rho_\popu$ rather than $\Sigma_{12,\popu}$. %
This is a notable result because one might expect a false correlation identified in the population distribution to align with the correlation in the parameter estimation space.
We see here that need not be the case.  

The variance $\sigma_{\rho_{\popu}}$ increases with both $\sigma_\pe$ and $|\rho_{\pe}|$.
This leads to a counterintuitive outcome: while wider or more strongly correlated single-event posteriors can lead to a larger shift in $\rho_{\popu,\ml}$, these effects also tend to produce wider credible intervals on the measurement, making it less likely that the $\rho_\popu =0$ case will be falsely ruled out, especially when $\sigma_\pe$ is large or $N$ is small.
This is demonstrated in the bottom panel of Fig. \ref{FIG:grid}.
We therefore infer that strong correlations at the single-event level could, in fact, obscure an existing population-level correlation simply by increasing measurement uncertainty.

On the other hand, increasing $N$ causes the spread of $\rho_{\popu,\ml}$ and the median and spread of $\sigma_{\rho_\popu}$ to shrink.
In other words, a larger catalog size means a measurement of $\rho_\popu$ that is both more accurate and more precise and, consequently, a decreased probability of measuring $\rho_\popu \neq 0.$ The one exception is when $|\rho_\pe|$ and $\sigma_\pe$ are largest, which leads to such great measurement uncertainty in the $N=50$ case that increasing the catalog size to $N=100$ does not lower the probability of a false-positive measurement. %\at{explain why}\cbo{Beyond the fact that uncertainy seems to grow faster in $\rho_\pe$ than it decreases in $N$?}

%-----------------------------------------------------------------------------------
\subsection{Mismodeling at the single-event level}\label{SUBSEC:BinaryMisModeling}
We move on to the case where the waveform used in single-event parameter estimation does not match the true waveform of nature.
However, we again consider a population distributed according to uncorrelated Gaussians with $\mu^{\star}_{\popu,1}=\mu^{\star}_{\popu,2} = 0$,  $\sigma^{\star}_{\popu,1}=\sigma^{\star}_{\popu,2} = 1$ and $\rho^{\star}_{\popu} =0$.
To emphasize the effects of mismodeling, we consider the case where $\sigma_{\pe,1} = \sigma_{\pe,2}=0.1$ and $\rho_\pe = 0.9$, because we saw in the previous section that, in the absence of waveform mismodeling, this case differed very little from that in which  $\rho_\pe = 0$. 

We consider several illustrative examples for $\Delta\boldtheta_\mathrm{sys}$.
We begin by considering the case where the bias depends only on $\boldtheta^{\star}$ and no additional parameters. We select the specific realization 
\begin{align}\label{EQ:bias_form}
        \Delta\theta_\mathrm{sys,i} = \varphi\, \theta^{\star}_{1} \theta^{\star}_{2}\,
\end{align}
for $i = 1,2$. The constant $\varphi$ parametrizes the size of the derivation and, for the time being, we take its value to be common among all events in a given catalog. We choose this realization for the bias because, while $\theta^{\star}_{1}$ and  $\theta^{\star}_{2}$ are drawn from uncorrelated distributions, $\Delta\theta_\mathrm{sys,1}$ and  $\Delta\theta_\mathrm{sys,2}$ will clearly be correlated across the catalog, inducing correlation in $\{\boldtheta^k_{\ml}\}$.

The statistical uncertainty in single-event parameter estimation scales inversely with the SNR, while the systematic bias remains constant. Therefore, as detectors improve and events are observed with greater sensitivity, the relative significance of systematic bias will grow.  
Significant systematic error is already present in some regions of parameter space \cite{LIGOScientific:2021usb,KAGRA:2021vkt,LIGOScientific:2025slb,LIGOScientific:2026wfs}.

We consider the systematic bias significant when the true source value of the parameter falls outside the 90\% credible bounds of the marginalized posterior. In other words 
\begin{align}\label{EQ:bias_ineq}
    |\Delta \theta_\mathrm{sys,i}| < \sqrt{2}\text{Erf}^{-1}(0.9) \sigma_{\pe,i}\,.
\end{align}
The single-event parameters $\theta_1$ and $\theta_2$ are unbounded, and the absolute value of the bias grows with distance from the $\theta_1=\theta_2=0$ point. Therefore, for any choice of $\varphi$, there is a region at the edges of the parameter space where the inequality above is violated. Thus, our chosen bias in Eq. \eqref{EQ:bias_form} is indicative of the current GW catalog.

\begin{figure}
    \centering
    \includegraphics[width=0.45\textwidth]{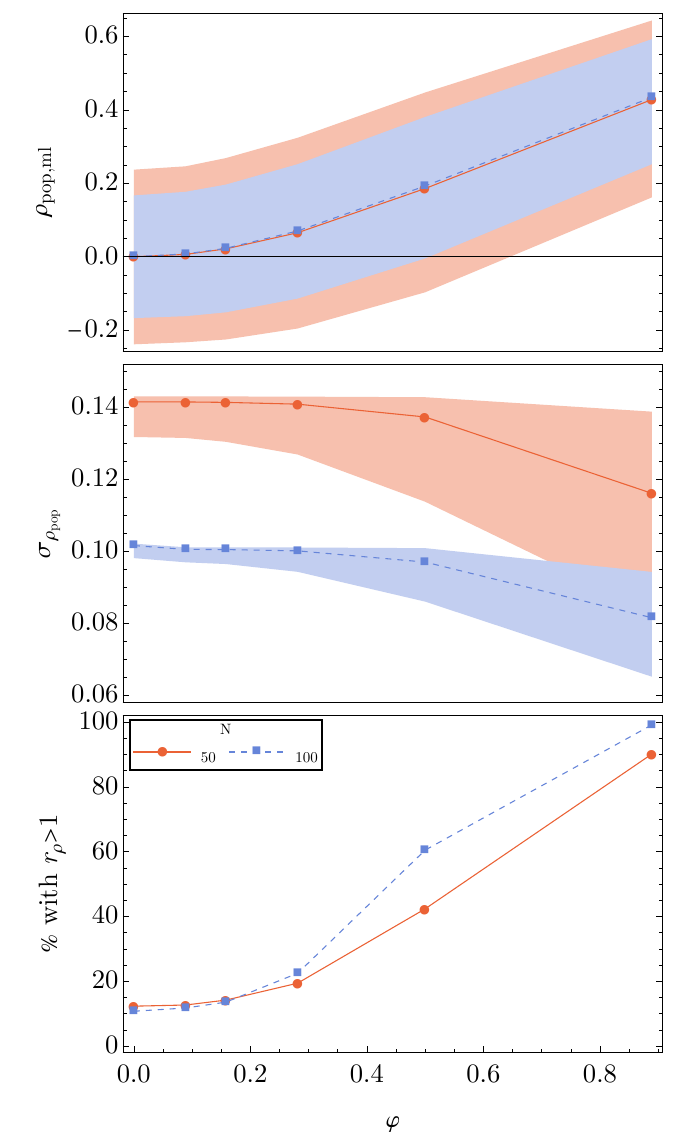}
    \caption{Top two panels show the median of the distributions $p(\rho_{\popu,\ml})$  and $p(\sigma_{\rho_\popu})$ as a function of $\varphi$ for catalogs of size $N$ with single-event likelihood speficied by $\sigma_\pe = 0.1$, $\rho=0.9$ and systematic error of the form in Eq.~\eqref{EQ:bias_form}. The $5\%$ and $95\%$ quantiles of the probability distributions enclose the shaded regions. The bottom panel shows the percentage of catalogs that produce a credible measurement of $\rho_\popu\neq0$, corresponding to $r_\rho>1$. The true population correlation is $\rho_\popu^{\star} =0$. The specific analyses considered here correspond to $\varphi = 5\times10^\beta$, for  $\beta = $ \{$-\infty$, $-1.75$, $-1.5$, $-1.25$, $-1.0$, $-0.75$\} which constitute systematic error that is significant for $\{0,~10,~20,~40,~50,~70\}\%$ of the catalog. We note that when systematic error is small, or nonexistent, increasing the size of the catalog lowers the probability of a false-positive measurement of $\rho_\popu\neq0$. However, as the size of the systematic error grows, the resulting increase in precision means a greater probability of obtaining a credible measurement of $\rho_\popu\neq0$.} 
    \label{FIG:percentvspercent}
\end{figure}

We consider $\varphi = 5\times10^\beta$, for  $\beta = $ \{$-\infty$, $-1.75$, $-1.5$, $-1.25$, $-1.0$, $-0.75$\}.
As a measure of the impact of the bias on the catalog,  for each value of $\varphi$ we compute the percent of the underlying population distribution for which the bias is significant according to the inequality in Eq.~\eqref{EQ:bias_ineq}.
We find that the above values of $\varphi$ correspond to systematic error that is significant for approximately $\{0,~10,~20,~40,~50,~70\}\%$ of the catalog, respectively.
While the largest of these percentages are greater than we suspect for our current GW catalog, by considering large biases, we are able to qualitatively explore the relationship between systematic error and catalog size in producing false-positive measurements of correlation. 

We carry out the procedure outlined in the previous section to approximate the distributions $p(\rho_{\popu,\ml})$, $p(\sigma_{\rho_\popu})$, and to determine the percent of catalogs for which $r_\rho>1$.
We consider catalogs with $N\in\{50,100\}$. Figure \ref{FIG:percentvspercent}
shows the median of the distributions $p(\rho_{\popu,\ml})$  and $p(\sigma_{\rho_\popu})$ as a function of $\varphi$.
The $5\%$ and $95\%$ quantiles of the probability distributions enclose the shaded regions.
The bottom panel shows the percentage of catalogs for which $r_\rho>1$.

We see that as the significance of single-event bias grows, the median $\rho_{\popu,\ml}$ moves farther from zero, indicating a higher probability for bias in the measurement of  $\rho_{\popu}$. While catalog size has virtually no impact on the median $p(\rho_{\popu,\ml})$, increasing the number of events does tighten the spread of the distribution.
On the other hand, as in the case with no mismodeling, catalog size is the dominant factor affecting the distribution $p({\sigma_\rho}_\popu)$, with larger catalogs corresponding to more precise measurements of $\rho_\popu$. 

When waveform mismodeling is small or nonexistent, the bias in $\rho_{\popu,\ml}$ is negligible, and a larger catalog size means a lower probability of measuring $\rho_\popu\neq0$.
However, as the waveform mismodeling and bias in $\rho_{\popu,\ml}$ grow, a larger catalog increases the probability of obtaining a credible measurement of $\rho_\popu\neq0$.
The effects of statistical uncertainty in the measurement of population parameters decrease as catalog size grows, allowing systematic effects to dominate.

Thus far, we have considered the case where $\varphi$ is common among all events. 
However, it is also possible that the bias in $\theta_1$ and $\theta_2$ depends on additional parameters that also vary across the catalog.
These parameters may be accounted for in the waveform approximant, or they may be additional, unmodeled parameters such as black hole hair or environmental effects.
We can represent such scenarios by allowing $\varphi$ to vary among events.

\begin{figure*}
    \centering
    \includegraphics[width=0.9\linewidth]{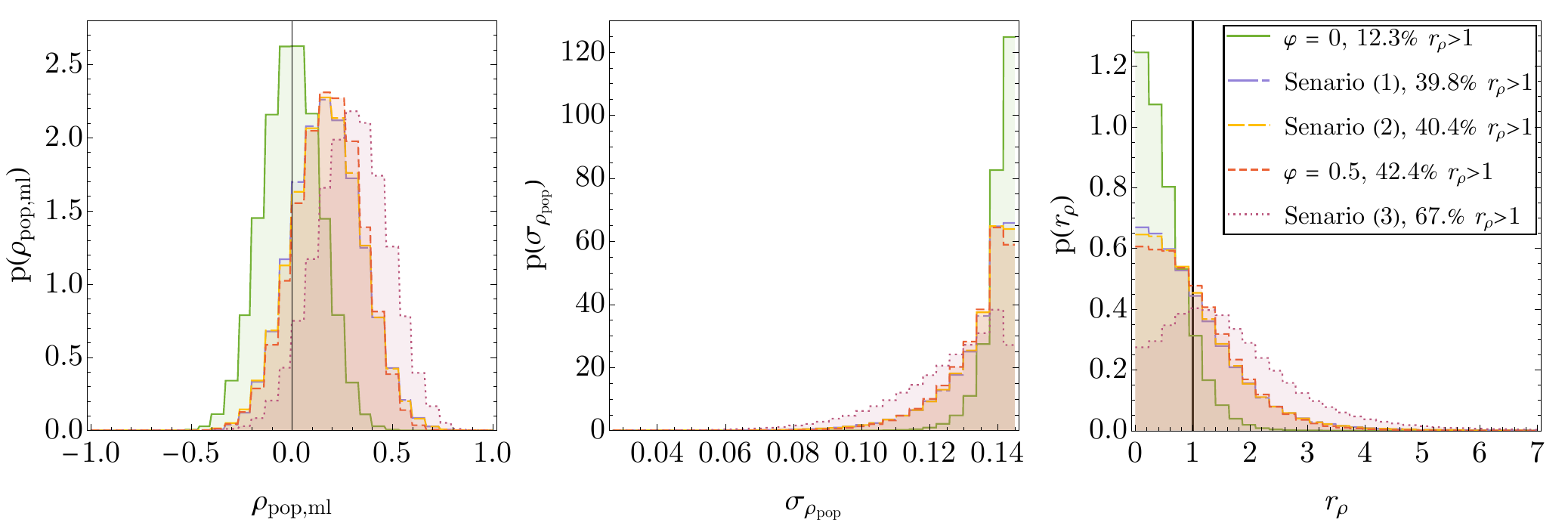}
    \caption{
    Probability distributions $p(\rho_{\popu,\ml})$, $p(\sigma_{\rho_{\popu}})$, and $p(r_\rho)$ for several scenarios in which systematic bias in single-event parameters takes the form given in Eq.~\eqref{EQ:bias_form}. In scenario (1),  $\varphi$ is drawn from a Gaussian distribution $\mathcal{N}(\varphi|0,0.5)$ centered at zero, in scenario (2), $\varphi$ is the absolute value of a number drawn from a Gaussian distribution $\mathcal{N}(\varphi|0,0.5)$ centered at zero, and in scenario (3), $\varphi$ is drawn from a Gaussian $\mathcal{N}(\varphi|0.05,0.5)$  with a nonzero mean. These senarios are compared to the case with no systematic error, $\varphi=0$, and the case where $\varphi=0.5$ is constant across events. Each distribution corresponds to a catalog of $N=50$ events, $\sigma_\pe=0.1$, and $\rho_\pe=0.9$. The black vertical line in the right-most panel indicates $r_\rho=1$, above which we assume a credible false-positive measurement of correlation. The percentage of catalogs in each configuration that fall above this line is reported in the plot legend. We see that a systematic error that depends on a third parameter can either increase or decrease the probability of a false-positive measurement of population-level correlation, depending on how that additional parameter varies across the parameter space.}
    \label{FIG:bias_plot}
\end{figure*}

We consider three specific scenarios: (1)  $\varphi$ is drawn from a Gaussian distribution $\mathcal{N}(\varphi|0,0.5)$ centered at zero, (2) $\varphi$ is the absolute value of a number drawn from a Gaussian distribution $\mathcal{N}(\varphi|0,0.5)$ centered at zero, and (3) $\varphi$ is drawn from a Gaussian $\mathcal{N}(\varphi|0.5,0.5)$  with a nonzero mean.
We compare these scenarios to the case with no bias ($\varphi=0$) and to the case where $\varphi$ is fixed ($\varphi=0.5$, corresponding to significant bias for 50\% of the population).
The distributions resulting from these scenarios are illustrated in Fig.~\ref{FIG:bias_plot}.
We see that allowing $\varphi$ to vary among events can either reduce or enhance the effects of systematic error with respect to the case of constant $\varphi$, depending on the distribution $\varphi$ takes.

%-----------------------------------------------------------------------------------
\subsection{Mismodeling at the population level}\label{SUBSEC:PopMismodeling}
Finally, we consider the case where the mismodeling occurs at the population level. 
That is, while we still employ the Gaussian model given in Eq. \eqref{EQ:GaussDists2} in the hierarchical inference, we draw the catalogs from underlying distributions $\pi^{\star}(\boldtheta|\boldLambda^\star)$ that are non-Gaussian.
We assume $\theta_1$ and $\theta_2$ are uncorrelated at the population level such that the joint probability distribution can be written as the product of the marginal distributions $\pi^{\star}(\boldtheta|\boldLambda^\star)=\pi_1^{\star}(\theta_1|\boldLambda_1^\star)\pi_2^{\star}(\theta_2|\boldLambda_2^\star)$.  
As with mismodeling at the single-event level, there are infinitely many ways in which the true population distribution might differ from the chosen model.
We focus on several underlying distributions that allow us to explore the effects of specific types of mismodeling. 

We consider two families of underlying distributions that generalize the Gaussian distribution. 
The exponential power distribution, 
\begin{align}\label{EQ:ExpPower}
   \pi_i(\theta_i|\sigma_{\popu,i},\kappa)\propto \exp\left({-\frac{1}{\kappa}\left|\frac{\theta_i}{\sigma_{\popu,i}}\right|^{\kappa }}\right)\,.
\end{align}
allows us to explore the effects of an underlying distribution with lighter or heavier tails than the model.
When $\kappa = 2$, we recover a Gaussian distribution. When  $\kappa>2$ ($\kappa<2$), the tails of the distribution are lighter (heavier) than those of a Gaussian distribution.
The skew normal distribution   
\begin{align}\label{EQ:Skew}
    \pi_i(\theta_i|\sigma_{\popu,i},\alpha)\propto&~\left[1+\mathrm{Erf}\left(\frac{\alpha}{\sqrt{2}}\frac{\theta_i}{\sigma_{\popu,i} }\right)\right]
    \eqbreak\times
    \exp\left[{-\frac{1}{2}\left(\frac{\theta_i}{\sigma_{\popu,i}}\right)^{2}}\right]\,.
\end{align}
allows us to explore the effects of using a symmetric model to recover an asymmetric distribution.
When $\alpha = 0$, we recover a Gaussian distribution. As the absolute value of $\alpha$ grows, the distribution becomes more asymmetric.
\begin{figure}[h]
    \centering
    \includegraphics[width=0.45\textwidth]{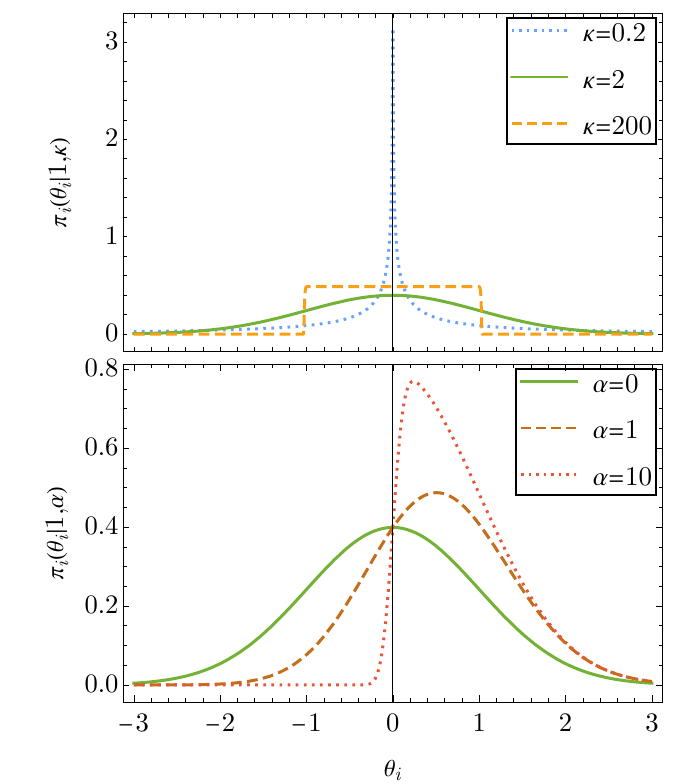}
    \caption{Marginal underlying population distributions used to generate the distributions in Fig. \ref{FIG:pop_mismodel}. The top panel depicts the exponential power distribution of Eq.~\eqref{EQ:ExpPower} and the bottom panel depicts the skew distribution of Eq.~\eqref{EQ:Skew}. 
    For each plot, $\sigma_{\popu,i} = 1$. 
    The green distributions ($\kappa=2$ and $\alpha=0$) are the same in both the upper and lower panels and correspond to a Gaussian underlying distribution.
    }
    \label{FIG:distributions}
\end{figure}

\begin{figure*}
    \centering
    \includegraphics[width=0.9\linewidth]{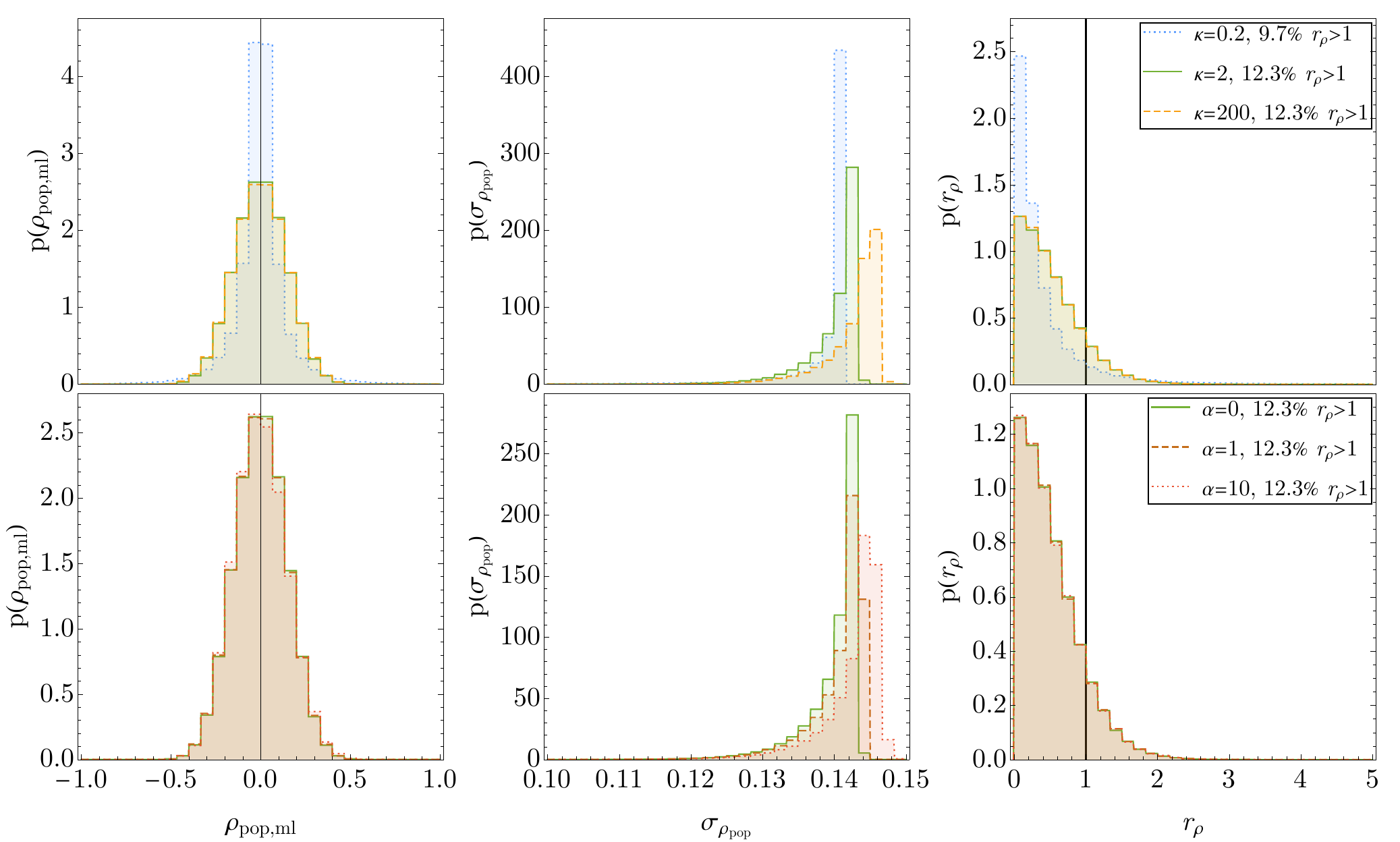}
    \caption{\label{FIG:pop_mismodel} Probability distributions $p(\rho_{\popu,\ml})$, $p(\sigma_{\rho_{\popu}})$, and $p(r_\rho)$  corresponding to several different uncorrelated underlying population distributions. In the top panels, the true population distribution is the exponential power distribution of Eq.~\eqref{EQ:ExpPower}, and in the bottom panels, the true population distribution is the skew normal distribution of Eq.~\eqref{EQ:Skew}. The green distributions ($\kappa=2$ and $\alpha=0$) are the same in both the upper and lower panels and correspond to a Gaussian underlying distribution. Each distribution corresponds to a catalog of $N=50$ events, a single-event likelihood specified by $\sigma_\pe=0.1$, and $\rho_\pe=0.9$, and the Gaussian population recovery model given in Eq.~\eqref{EQ:GaussDists2}. The black vertical line in the right-most panel indicates $r_\rho=1$, above which we assume a credible false-positive measurement of correlation. The percentage of catalogs in each configuration that fall above this line is reported in the plot legend. We see that for this simplified toy model, population mismodeling has a negligible effect on the probability of obtaining a credible measurement of the population-level correlation. The one exception is the exponential Gaussian population distribution with $\kappa = 0.2$, corresponding to a heavier-tailed distribution, which shows a reduced probability of a nonzero correlation measurement.}
\end{figure*}

We consider four underlying population distributions, each with $\sigma^{\star}_{\popu,1} =\sigma^{\star}_{\popu,2} =1$.
Two populations are distributed according to joint exponential power distributions with $\kappa^{\star} \in \{ 0.2 , 200\}$.
Two are distributed according to joint skew normal distributions with $\alpha^{\star}  \in \{ 1,10\}$.
For illustration, the marginal distributions are depicted in Fig. \ref{FIG:distributions}, compared to a Gaussian distribution (corresponding to an exponential power distribution with $\kappa=2$ or a skew normal distribution with $\alpha=0$).
We specify the single-event likelihood with $\sigma_\pe =0.1$ and $\rho_\pe=0.9$ and $\Delta\boldtheta_{\mathrm{sys}}=0$. For each underlying population distribution, we consider catalogs with $N=50$.
The probability distributions $p(\rho_{\popu,\ml})$, $p(\sigma_{\rho_\popu})$ and $p(r_\rho)$ are shown in Fig. \ref{FIG:pop_mismodel} along with those generated in Sec.~\ref{SUBSEC:NoMismodeling}, from an underlying Gaussian distribution. 

As discussed in Sec. \ref{SUBSEC:HI}, we do not expect $\rho_{\popu,\ml}$ to be biased from the true value $\rho_\popu^{\star} =0$ in the large $N$ limit, and this is indeed what we see, even for $N=50$.
On the other hand, the distribution $p(\sigma_{\rho_\popu})$ varies across underlying distributions, indicating that using an inaccurate population model could affect the precision with which correlation is measured. 
However, for most of the population distributions considered, we observe no differences in the percentage of catalogs that ultimately yield a false-positive correlation measurement compared with the case of no population mismodeling.
The one exception is when the population is distributed according to an exponential power distribution with a heavy tail.
In this case, though we generally associate mismodeling with a higher probability of bias, $p(\rho_{\popu,\ml})$ has smaller variance about the true value $\rho_\popu^{\star}=0$, resulting in a reduced probability of making a false-positive measurement of population-level correlation.

We have discussed how the simplicity of our Gaussian toy model makes it ill-suited to fully exploring the effects of systematic bias arising from population-level mismodeling.
To compound the issue, the population models used in the analysis of the actual GW transient catalog are more complex than the simple correlated Gaussians considered here, so mismodeling is likely to have effects that are not straightforward to predict.
However, our results suggest that while population mismodeling can affect the probability of a false-positive measurement of population-level correlation, an otherwise poor model may still identify the presence or absence of correlation in a population.

%%%%%%%%%%%%%%%%%%%%%%%%%%%%%%%%%%%%%%%%%%%%%%%%%%%%%%%%%%%%%%%%%%%%%%%%%%%%%%%%%%%%
\section{Conclusion and Discussion}\label{SEC:Conclusion}

We have used a toy model to explore how statistical and systematic effects at both the single-event and population levels may affect the measurement of population-level correlations between pairs of binary parameters in compact binaries observed by LVK.
The simplified nature of the toy model makes it difficult to draw quantitative comparisons with population analyses of the actual GW catalog.
That being said, this study allows us to draw several qualitative conclusions that may inform our understanding of the robustness of the identified population-level correlation.

We tackled the effects of single-event correlations on the measurement of population-level correlations, and we found two notable and counterintuitive results: 
\begin{itemize}
\item [(i)] While the presence of single-event correlation tends to increase the magnitude of the statistical shift in the maximum likelihood population correlation, this shift need not align with the correlation in parameter estimation.
Rather, there seems to be a slight negative correlation between single-event correlation and inferred population correlation when quantifying correlation with the dimensionless coefficient $\rho_\popu$. 
\item [(ii)] Because the presence of single-event correlation increases the uncertainty in the measurement of population correlations, it can actually reduce the probability of a credible false-positive correlation measurement at the population level.
We conclude that, for pairs of parameters with strong correlation at the parameter estimation level, such as effective spin and mass ratio, this correlation could potentially obscure true underlying population correlations.
This effect will diminish as the SNRs of detections increase and as more events are added to the catalog. 
\end{itemize}

We next considered systematic effects resulting from mismodeling in both the waveform and population distribution. For the latter, we found evidence that an otherwise poor population model may still be useful for ruling out correlation in the underlying distribution. As for waveform systematics, we found that when waveform mismodeling introduces biases into recovered single-event parameters that are correlated across the catalog, these biases are likely to be interpreted as population-level correlations between those parameters.
Unlike the effects of single-event correlation, this effect will tend to increase both as detections become louder and as more events are added to the catalog.
Our results align with general intuition about systematic effects: increasing detector sensitivity and adding more events to the catalog will make our measurements of single-event and population parameters more precise, but can only make them more accurate to the extent that our waveforms and population models are good representations of reality. 

The state-of-the-art approximants \texttt{IMRPhenomXPHM-} \texttt{SpinTaylor} \cite{Pratten:2020ceb, Colleoni:2024knd},  \texttt{SEOBNRv5PHM} \cite{Ramos-Buades:2023ehm,Pompili:2023tna} and \texttt{NRSur7dq4} \cite{Varma:2019csw} 
used to analyze the current GWTC-5 events \cite{LIGOScientific:2026wfs} are meticulously validated against numerical-relativity simulations.
However, the former approaches recover numerical-relativity waveforms less faithfully in regions with more unequal mass ratios and higher spin precession, while the latter can only capture sufficiently short signals. This mismodeling, which varies across the parameter space, has the potential to produce spurious population-level correlations.
Several events in the current GW catalog exhibit discrepancies between the posteriors obtained using different waveforms, indicating the presence of systematic error. 
Unsurprisingly, this error is often most evident in the mass ratio and spin parameters~\cite{LIGOScientific:2021usb,KAGRA:2021vkt,LIGOScientific:2025slb,LIGOScientific:2026wfs}.
Our results emphasize the importance of investigating how such biases could be affecting correlation measurements in the actual GW catalog.

Furthermore, we have focused here on systematic error arising from mismodeling in the waveform approximant and the population distribution model.
However, systematic error is, more generally, any non-stochastic error arising from incorrect assumptions about the experiment.
This could include, among other things, deviations from the assumption of stationary and Gaussian noise that inform the choice of single-event likelihood or mischaracterization of selection effects at the population level.
While we did not consider such effects in our study, they could potentially contribute to biases that could be interpreted as population correlation. 

As the number of detected GW events grows well into the hundreds, the marginal population distributions are becoming increasingly well understood, and identifying correlations and subpopulations is emerging as a prominent new frontier for characterizing the observed dataset and ultimately improving our (astro)physical understanding of compact binaries.
Despite the simplicity of our model, we hope this exploration of the robustness of inferred correlations will serve as a useful compass for interpreting current and upcoming GW data releases.

\acknowledgements
We thank Riccardo Sturani and Matthew Mould for discussions.
C.B.O., A.T., and D.G are supported by 
supported by 
ERC Starting Grant No.~945155--GWmining, 
Cariplo Foundation Grant No.~2021-0555, 
Italian-French University (UIF/UFI) Grant No.~2025-C3-386,
MUR Grant ``Progetto Dipartimenti di Eccellenza 2023-2027'' (BiCoQ),
and the INFN TEONGRAV initiative.  
A.T. and D.G. are supported by MUR Young Researchers Grant No. SOE2024-0000125.
D.G. is supported by MSCA Fellowship No.~101149270--ProtoBH.
Computational work was performed 
at CINECA with allocations through INFN and the University of Milano-Bicocca.

\end{document}